# MPSM: Multi-prospective PaaS Security Model

Robail Yasrab
*Departmnet of Computer Science and Technology*
*University of Science and Technology of China*
*Hefei, China*
robail@mail.ustc.edu.cn

*Abstract*—Cloud computing has brought a revolution in the field of information technology and improving the efficiency of computational resources. It offers computing as a service enabling huge cost and resource efficiency. Despite its advantages, certain security issues still hinder organizations and enterprises from it being adopted. This study mainly focused on the security of Platform-as-a-Service (PaaS) as well as the most critical security issues that were documented regarding PaaS infrastructure. The prime outcome of this study was a security model proposed to mitigate security vulnerabilities of PaaS. This security model consists of a number of tools, techniques and guidelines to mitigate and neutralize security issues of PaaS. The security vulnerabilities along with mitigation strategies were discussed to offer a deep insight into PaaS security for both vendor and client that may facilitate future design to implement secure PaaS platforms.

*Keywords*-PaaS; SaaS; IaaS; Security; Privacy; Virtual Machine (VM); Container

## I. INTRODUCTION

These days, organization and businesses are considering cloud computing as key technology architecture to increase efficiency and save money (Chou, 2015). However, most organizations are currently concerned about the cloud associated security aspects. Most of the security issues emerge due to the shared nature of cloud. Though, cloud computing is not basically an insecure technology infrastructure; it just requires secure access and management.

Munir et al. (Munir and Palaniappan, 2013) stated that through eliminating security issues, it is likely to enhance the cloud computing efficiency and acceptance. Higher acceptance of cloud technology will lead to better technology resource utilization, enhance technology adaptation, ease new services deployment, facilities maintenance and decreases expenditures. Cloud Computing has broadened its scope to new technology paradigms; it is categorized into three technology models. Software-as-a-Service (SaaS) offers ready-to-use software services. Infrastructure-as-a-Service (IaaS) provides virtual computers for clients. Finally, Platform-as-a-Service (PaaS) offers a homogeneous computing platform to develop and run applications. Microsoft Azure and Google App Engine (GAE) are popular examples of PaaS model. PaaS offers capabilities to develop and implement custom applications. It provides an online computing platform as a service. According to Devi et al. (Devi and Ganesan, 2015) PaaS enables development of apps without the complexities of purchasing and managing the underlying software and hardware. It offers all of the facilities necessary for whole software development lifecycle. Currently, online cloud based PaaS platforms are offering tools and support for application design, development, testing, deployment and hosting. It includes a complete application services package that comprises web service integration hosting services, team collaboration, database integration, resource scalability, storage management, security and developer community facilitation.

Sandikkayavi et al. (Sandikkaya and Harmanci, 2012) stated that PaaS shared development environment presents some unique security challenges like access control, authentication and authorization. It's distributed storage and applications nature also involves data security issues. There is always danger of mission critical information theft that can be obtained during a data breach. Hackers are always looking for security and configuration loophole to attack and destroy intellectual resources. PaaS security is not solely the vendor responsibility. A number of security aspects need to be analyzed before choosing a PaaS vendor. This research aimed to asses and analyzes the security issues of PaaS cloud infrastructure. The prime objective of this study was to outline most vulnerable security aspects and their impact. This research has recommended a PaaS security model for better handling and management of security issues in cloud infrastructure. This model was a multi-prospective model that was aimed to address different levels of PaaS architecture.

PaaS technology basic architecture presented in Section I. Section II comprised the details of the security issues and problems to the PaaS platform. In section III presented the proposed security model for PaaS cloud. Section IV reserved for discussion and analysis. This section also outlined the future research directions. Conclusion and references were presented in last section.

## II. ARCHITECTURE OF PASS

PaaS architecture is based on two fundamental layers: Communication Layer and Management Layer. Availability,

reliability and optimization are key features those are monitored and handled by the communication layer for running applications on cloud. The Management Layer deals with provisioning and management capabilities. It is responsible for catering the scalability and reporting requirements of the custom PaaS (Jaiswar, 2015). Fig. 1 shows the overall architecture of PaaS platform.

*A. Communication Layer*

Communication Layer contains components/ services that deal with communication between the outer world and business services/ components. Communication Layer tools guarantee that the applications are failure-resistant. It also optimizes the execution of applications. Key components of communication layer are:

Load Balancer: This module deal with all executing application services and keep track requests being serviced by each app. This module maintains reliability, availability and optimal usage feature of the services.

Cloud Service interaction: Cloud services might move from one VM to another so they do not have a permanent address. Cloud Service interaction module guarantees the reliable interface with these services, thus ensuring availability.

Cloud Service Orchestration: This module ensures that each execution is well-maintained and reused afterward when the process has to be restarted in the case of failure. Cloud Service Orchestration ensures the optimal usage of technology resources that is one of the key requirements of PaaS technology architecture (Jaiswar, 2015).

Caching: Caching is simply a common cache for the whole cloud architecture. It is different from traditional cache as it may not reside on a particular virtual machine. It is hard to ensure the utmost reliability of the cache.

Messaging: The messaging module is extremely reliable component. It can be operated from multiple VMs. In case, if some VM fails, messaging module ensures that there will be no single point of failure. It also ensures that services can easily communicate with each other. Virtual Machine (VM) Management: This module offers interaction among the primary virtualization solution and the application services. For example, it can be used by the load balancer to query a virtual machine for its CPU utilization. It is also capable to offer interface among multiple virtualization engines.

*B. Management Layer*

This layer offers management and monitoring services in the PaaS architecture. Some of the key modules of this layer are: Provisioning: This module offers provisions application services. These services originate from on-demand request formulated through the portal. In PaaS architecture integrated provisioning component may be formulated through coupling provisioning engine and VM provisioning engine.

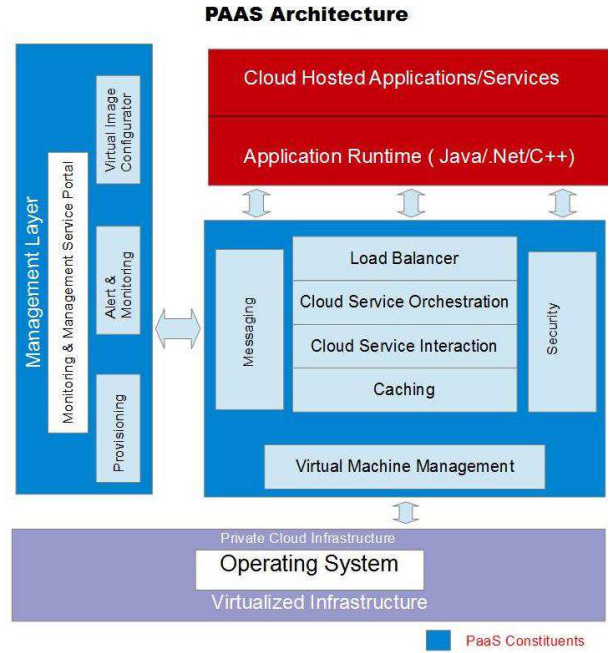

Figure 1. Architecture of PaaS

Alerts & Monitoring: Monitoring the application execution services and alerting any failures is the key responsibility of this module. This alerting and monitoring module can be constituent of the organizational monitoring tool for the cloud platform. Image Configuration: This module acts as a reliable cloud store that permits storage of application images as blobs. It also offers capability to create new images and their mapping to VM profiles.

Management & Monitoring Service Portal: This module is a dashboard for application services monitoring. It also offers a user interface for provisioning new services.

III. REQUIREMENTS OF A PAAS PLATFORM

PaaS is a runtime environment for cloud applications. One of the key objectives of PaaS technology is to enhance the cloud efficiency and improve its benefits. From application development prospective an ideal PaaS platform must fulfill following requirements (Jaiswar, 2015): High Scalability and On Demand Provisioning: An ideal PaaS expected to stretch apps to the last hardware resource available for deployment. This offers to user a feeling of infinite scalability. Moreover, the application provisioning should be automated. There should be no IT interference for apps deployment and delivery. High Availability: A high quality PaaS environment should offer high availability capabilities. It should exhibit excellent support for applications development process in case of any disaster or failure. Features like failover and load balancing are one of the key capabilities required.

High Reliability: Reliability and availability are often used interchangeably. These both capabilities are intended to offer support for failover. A reliable PaaS system should respond within the specified time. An efficient PaaS platform should offer reliability to the entire services/components running under them.

Optimal Usage: Optimal usage of resources is one of the key requirements of any cloud computing platform. PaaS environment demands highly optimal usage of resources for executing applications. A high quality PaaS platform should have tools and services to monitor application execution and usage to attain better resource optimization. Auto-Scaling: Auto-Scaling is the one of the key feature of any cloud service. A good PaaS service provider should offer auto-scaling to the applications in case of newly added computing resources. It shows an elastic nature of PaaS platform that can be expanded and add more resources according to need of users.

Admin /Management Console and Reports: Technology resources need to be tracked and monitored through a user interface in a high quality PaaS platform. PaaS platforms should offer reporting tools to outline statistics regarding application execution, usage and provisioning. Multi OS and Multi Language Support: An ideal PaaS should offer wide ranging support to multi operating systems and multi programming languages. It should support runtimes which can execute on multiple operating systems (Linux, Windows etc.) and should support different programming languages (PHP, Java, .Net, C++ etc.).

## IV. SECURITY AND PRIVACY RISK UNIQUE TO PASS ENVIRONMENT

The unique PaaS environment raises several security and privacy concerns. Security issues related to the Service oriented-architecture (SOA) are one of the key concerns in PaaS environment. As PaaS is based on the SOA model, therefore, SOA domain related security issues inheriting to PaaS architecture. Almorsy et al. (Almorsy et al., 2010) outlined that PaaS is currently facing security and privacy issues like man-in-the-middle attacks, replay attacks, XML-related attacks, injection attacks, input validation related attacks and dictionary attacks. These are security and privacy risks emerged due to unique nature of PaaS architecture. This unique cloud architecture makes PaaS more vulnerable for outsider and insider attacks. The following features will show how PaaS cloud nature makes it more vulnerable for security and privacy attacks.

Virtualization: communications between virtual machines can be vulnerable to security and privacy attacks. There needs ensuring mediated sharing, strong isolation and communications between VMs.

Outsourcing: PaaS cloud users may lose control of their data.

Multi-tenancy: Cloud is all about multi-tenancy. Sharing resources among processes can be dangerous. PaaS secure multi-tenant environment demands access policies, data access protection and secure application deployment.

Heterogeneity: Different PaaS providers may have different tools and services to provide security and privacy mechanisms. This could lead to integration challenges (Shahzad, 2014).

Extensibility and Shared Responsibility: Different delivery models exhibit different level of security responsibility for service providers and customers.

There is need for effective authorization, mutual authentication and Web Services (WS) security standards to secure the PaaS cloud provided services. The security aspect is a mutual responsibility among PaaS service providers, cloud providers and user.

## V. PAAS SECURITY CHALLENGES

Figure 2. shows all possible security challenges to PaaS infrastructure.

### A. General Cloud Security and Privacy Challenges

**Security Issues**: - Security is protection of valuable and sensitive data of client from vulnerable attacks. Numerous risk factors are outline below regarding security issues in cloud computing environment.

A. Multitenancy: Multitenancy allows one application to run on multiple systems at the same time however it can causes security vulnerabilities in cloud computing infrastructure (Munir and Palaniappan, 2012) (Rai et al.) (Mahmood, 2011) .

B. Access: The intellectual resources and sensitive data of client have many security threats. Any security breach can cause huge damage to business or individual (Munir and Palaniappan, 2012) (Mahmood, 2011) (Attas and Batrafi, 2011).

C. Availability: Availability is one of the key concerns in cloud infrastructure. Cloud resources need to be available to user at anytime and anywhere. Cloud downtimes or backup/recovery issues are one of key aspects that cloud users concerned today (Attas and Batrafi, 2011) (Krutz and Vines, 2010).

D. Trust: Cloud consumers cannot completely trust on the service providers about their sensitive data to be store on cloud. Cloud service providers are still not informing about the location of client's data. Therefore, there is a trust gap between client and cloud vendors (Krutz and Vines, 2010) (Sharma et al., 2011).

E. Audit: Cloud Solution Provider (CSP) needs to perform internal monitoring control through external audit mechanism. It will ensure high quality security capability demonstrated by service provider. However, most of CSP still fails to offer fair audit mechanism with effecting integrity (Krutz and Vines, 2010) (Shaikh and Haider, 2011).

**Privacy Issues**: - danger of leakage or misuse of sensitive data of customer. Data can be breached while transmission or through some passive attacks.

A. Misuse of Cloud computing: Sometimes unlimited access of storage or network resources offered by cloud services provider may cause different harms to cloud platform. Free trials offered by service providers also one of the key cause of misuse of cloud resources (Mather et al., 2009) (AlZain et al., 2012).

B. Malicious Insiders: Some unsatisfied employee can cause huge damage to corporate resources. Sometimes, attackers get access to employee details to perform a malicious attack (Munir and Palaniappan, 2012) (Munir and Palaniappan, 2013) (Sravani and Nivedita).

C. Trans-border data flow and data proliferation: These days data proliferation becomes a big problem for cloud service providers. Management and transfer of huge amount of data is a big concern. Currently, it is hard to ensure that duplicates of the user's sensitive data or its backups are not processed or stored anywhere. It is also hard to guarantee that all copies of data are deleted, if a delete request is made (Mather et al., 2009) (AlZain et al., 2012) .

D. Dynamic provision: It is hard to ensure that the secrecy of user's private data in the cloud. No one is taking the responsibility of security of the consumer data. Dynamic provision is an effective way to ensure security of data in cloud (Mather et al., 2009) (AlZain et al., 2012) (Chen and Zhao, 2012).

| Cloud Type | Category | Challenges |
|---|---|---|
| General Cloud Computing security issues | Privacy Issues | Misuse of Cloud Computing -Malicious Insiders -Trans border data flow and data proliferation -Dynamic provision |
| | Security Issues | -Audit -Trust -Availability -Multitenancy -Access |
| PaaS specific Cloud security issues | Data and Infrastructure related security issues | -Data location -Information leakage -Mashups -Privileged access -Distributed system -Heterogeneous Technology Resources -Infrastructure security -Vulnerable host |
| | Security Attacks | -Denial of Service (Dos) Attacks -Man-in-the-Middle (MitM) Attacks -ARP spoofing -Poisoned Images -Compromised Secrets |

Figure 2. PaaS Cloud Challenges

*B. PAAS Level Security Challenges*

A. Data location: PaaS service provider duplicate data on multiple locations to offer a high availability of data for clients. Even user delete the data, it's never completely deleted. Instead the pointers are deleted those pointing to data. The data remains on service provider's network and its exact location is unknown. This is a big security issues that still threaten PaaS users (Subashini and Kavitha, 2011) (Almorsy et al., 2010).

B. Privileged access: PaaS platforms offer a popular feature of "built-in debug". Developers usually use this feature to discover problems in code. Debugging features grant privileged access to memory and data locations (Subashini and Kavitha, 2011) (Almorsy et al., 2010). It allows stepping through code and altering values to check different results. Debugging features grant highly privileged access to technology resources that is good for developers. Though, hackers can use this tool to exploit computing resources or to perform malicious attacks.

C. Distributed system: File system at PaaS is often highly distributed. Hadoop distributed file system (HDFS) is one of the popular implementation. HDFS uses 50070, 50075 and 50090 as default ports (TCP ports). Sample (Sample, 2015) states that these ports represent attack vectors where a hacker can attempt a numerous inputs to cause Denial of Service (DoS) attacks or system failures.

D. Heterogeneous Technology Resources: Heterogeneous technology resources (hardware and software) are unified in PaaS architecture for efficient use. Though, heterogeneity may cause some security settings issues due to diverse computational resources (Rocha and Correia, 2011). Security breaches can occur if objects' access to the technology resources cannot be controlled in a standard way.

E. Information leakage: Information leakage is one of the big problems in PaaS cloud. It could occur due to shared communication channels and resources. In PaaS distributed architecture, less secure communication channels can cause information leakage (Ristenpart et al., 2009)

F. Vulnerable hosts: Multi-tenancy allows user objects to communicate over interconnected multi-user hosts. Malicious objects and hosts can cause possible attacks in a multi-tenant environment. An attacker can access the host's resources and its entire tenant objects by breaching host security (Rocha and Correia, 2011) (Van Dijk and Juels, 2010) (Sandikkaya and Harmanci, 2012)

G. Mashups: Currently, PaaS not only offers traditional programming languages support, however it also offers third-party web-services components (e.g. mashups). Mashups are combination of more than one source element. Mashups could present data and network security threat in PaaS model. These are third-party services, so developer need to be aware while incorporating these services (Hashizume et al., 2013).

H. Infrastructure security: Infrastructure security is a big aspect in PaaS cloud. Generally, developers do not have access to the underlying PaaS infrastructure layers, so cloud vendors are responsible for infrastructure security. There is no guarantee that tools and services provided by vendor are secure or not (Hashizume et al., 2013).

I. Denial of Service (Dos) Attacks: Denial-of-Service attacks (DoS attacks) are well-known attacks on technology networks. In the PaaS architecture all resources are shared (Bakshi and Yogesh, 2010). DoS attack condition happens when one resource/user/object exploits access to a resource. In such conditions, it gets control of all other resources/users/objects (starvation).

J. Poisoned Images: You et al. (You et al., 2012) pointed out that at PaaS Platform, images may be injected through some viruses or Trojan infected software. Furthermore, problem of poisoned images may also be happened if someone is running outdated and known-vulnerable software versions. Such poisoned images could destroy valuable data or cause data leakage.

K. Compromised Secrets: Compromising business or personal secrets are the big security dangers in container based technology. In case of theft of the API keys and database passwords; the overall system can be compromised (You et al., 2012). Cloud technology is so vulnerable regarding compromising business and personal secretes.

L. Man-in-the-Middle (MitM): In such attacks a malicious actor inserts himself/herself into a communication among two legitimate parties at PaaS platform (Subashini and Kavitha, 2011). Such attacker monitor, alter or steal valuable information which is being transmitted among two parties.

M. ARP spoofing: ARP (Address Resolution Protocol) spoofing is a kind of security attack in which an attacker sends fake ARP messages over a LAN (local area network). WU et al. (Wu et al., 2010) stated that in such attacks, attacker is able to link his/her MAC address with the IP address of a legitimate system on the network. As the attacker's MAC address is linked to legitimate IP address, so the attacker will start getting each and every bit of data from that specific IP address. In PaaS, if an attacker able to gains access to one of the user's platforms, through compromising the security, the attacker can obtain, manipulate or redirect any information of the bridge. This information can be of any kind of data which could be transmitted among PaaS cloud and the outside world.

## VI. PROPOSED SOLUTION: MULTI-PROSPECTIVE PASS SECURITY MODEL (MPSM)

This research has proposed a multi-prospective PaaS Security Model (MPSM). It is based on set of guidelines for improving the security in PaaS infrastructure. This model composed of three major aspects: PaaS components, restriction level and security model. Fig. 3 shows a cube that demonstrates the MPSM. It is a layered security model

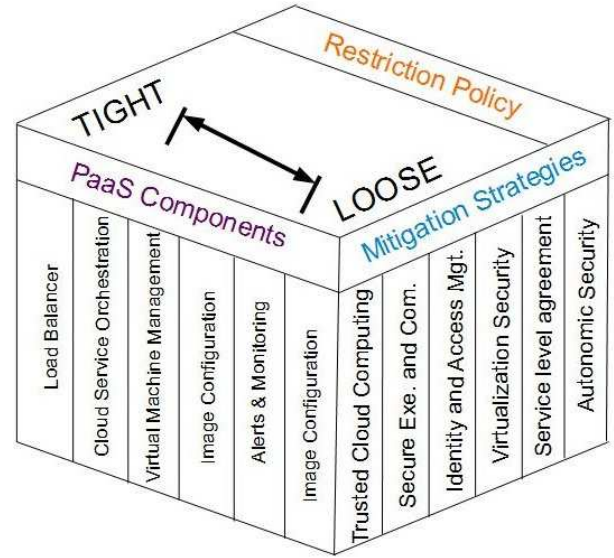

Figure 3. Multi-prospective PaaS Security Model (MPSM)

for data security that implements tools in every layer of the PaaS cloud architecture.

This model includes the key components of PaaS as shown in the front side of cube. These components are discussed in earlier section. The mitigation strategy side (2nd vertical side) of cube comprises the key security management components. It shows the security management policy, tools and techniques. This side of cube is basically denotes the mitigations for aforementioned security vulnerabilities. Methods applied are trusted cloud computing, identity and access management (IAM), virtualization security management, service level agreements for cloud security and autonomic security. These policies, tools and techniques are intended to offer a multi-dimensional security for the PaaS architecture. Further details of these components are outlined below. The third side of cube is restriction policy. It denotes the security model entities' restriction level. It is an adjustable mechanism that can be adopted according to the requirements of PaaS service provider and clients. This whole model shows the relationship among the security requirements and PaaS components. It shows how PaaS security infrastructure can be fully secured through adopting more enhanced services and application of different security tools.

### A. Trusted Cloud Computing

Trusted PaaS cloud architecture is intended to protect the cloud systems from malicious intrusion and security breaches. It ensures that PaaS cloud technology resources will act in a safe and predictable manner. Trusted cloud computing based on following key practices:

- Hypervisor and Application data protection
- Restricting illegal access to sensitive information

- Storage Authentication
- Encrypting devices to protect data in case devices are stolen or lost
- Hardware and software mechanisms compliance

Trusted cloud computing (TCC) is based on trusted cloud base (TCB) that is combination of trusted hardware, software and firmware to reinforce the security policy. TCC implements a security boundary that separates TCB from rest of the system. TCC also enforces a trusted path between user and systems to perform necessary operations.

*B. Secure Execution and Communication Environment*

PaaS is a multi-tenant environment, where everybody is security concerned. Service providers should take care of physical security, which is an important aspect of overall cloud security architecture. Application and network-level security is the responsibility of the user. Execution and communication security is one of major concern for PaaS service providers. National Institute of Standards and Technology (NIST) specified confidentiality, integrity and availability guidelines should be fulfilled to ensure PaaS cloud security. PaaS cloud confidentiality could be achieved through applying network security protocols, authentication services and encryption tools. Integrity could be ensued through implementing network firewall services and intrusion detection systems. Finally, availability demands fault tolerance architecture with backup facility and redundant disk management systems.

*Security management tools and techniques*: PaaS cloud comprises three types of data i.e. transmission data, storage data and processing data. There is security management tool for every kind of data as shown in Table 1.

Table I
DATA SECURITY MANAGEMENT TOOLS

| Storage | Processing | Transmission |
|---|---|---|
| Symmetric encryption | Homomophric encryption | Secret socket layer SSL encryption |
| AES (Advanced Encryption Standard) is a symmetric encryption algorithm | Partially homomorphic cryptosystems ElGamal cryptosystem | Transport Layer Security (TLS) 1.3 (Bhargavan et al., 2015) |

A. Protection of data in transmission: Protect data in transit is a really important especially when data is sensitive or confidential. Transport Layer Security (TLS) 1.3 is the latest standard for secret socket layer (SSL) encryption. It offers a secure connection because transmission is protected through symmetric cryptography. For each connection, keys are generated uniquely through secret negotiation (Handshake Protocol).

B. Protection of stored data: To protect sensitive and confidential storage data in PaaS cloud, Advanced Encryption Standard (AES) is the best security encryption algorithm. U.S. National Institute of Standards and Technology (NIST) specify AES for the encryption of electronic data. AES is based on symmetric-key algorithm, because it uses same key for both decrypting and encrypting. ISO/IEC 18033-3 standard also included AES as key security management tool.

C. Protection of data in Processing: ElGamal is an asymmetric key encryption algorithm. It is based on the "Diffie-Hellman key exchange". ElGamal cryptosystem is a partially homomorphic cryptosystem that can be used for protection of data during processing at PaaS cloud. ElGamal cryptosystem allows computations to be carried out on cipher-text. The results generated are encrypted, so it is the key advantage of ElGamal. It allows communication among services without exposing the data to each other.

*C. Identity and Access Management (IAM)*

Identity and Access Management (IAM) can be described as offering an adequate level of security for organizational resources and data. IAM employs policies and rules which could be applied on users via different methods such as allocating privileges to the users, enforcing login/password and provisioning user accounts.

A. Authentication: Authentication is one of the key elements of access control. It enforces parties to verify the authenticity during an interaction. Confidentiality of communication and identity authentication requirements can be attained by employing Transport Layer Security (TLS) where a public key infrastructure (PKI) exists. Certificate-based authorization is one the strong authentication method. PKI facility module issue certificates those can be used for administering access control. Extended X.509 certificate is an example of strong certificate-based authorization that carries user's role information (Zissis and Lekkas, 2012). It is really significant to encapsulated access control policies with corresponding objects. It offers high authentication, object scale customization and ease of distribution.

B. Authorization: It is the second key element of access control procedure. Authorization methods decide who can access to the objects based on pre-established procedures. Policy enforcement points (PEPs) can be used to attain high authorization (Zissis and Lekkas, 2012). PEP deals with the objects access, according to the policy defined inside the objects. PEPs of the hosts are responsible for evaluating the decision for authentication of any object.

C. Traceability: It is third and last element of access control. It is applied through keeping track records of the key events occurred in a system. It is also used to assess service characteristic by evaluating events record (Nhlabatsi et al., 2014). To implement high traceability at PaaS cloud there is need for an integrated undeniable logging method. This logging method must be secured and protected from all interacting clients (also system administrators).

*D. Virtualization Security Management*

A. Harding the Virtual Machine (VM): VM need to be configured according to vendor provided security guidelines. Resource consumptions, network interface and storage configuration should be performed carefully (Krutz and Vines, 2010). Components might be shared across virtual network devices should be isolated and secured through appropriate security policy.

B. Harden the Hypervisor: Hypervisor is considered to be a critical focus as an attack vector. During adopting PaaS technology, it is critically important to ensure that hypervisor is deployed securely (Krutz and Vines, 2010). There is a great deal of need to employ change and configuration controls to hypervisor. It is standard and best practice to engage third party testing services to ensure the hypervisor strength against published vulnerabilities.

C. Firewall additional VM ports: Minimum level of access control should be given to the host systems by independently firewalling them. Any open multiple ports connecting to host system should be firewalled.

D. Root Securing Monitor: In VM based architecture, to defend against privileged escalation security threats, there is need to "root secure" the monitor (Krutz and Vines, 2010). In this way, any virtualized guest will lose all privileges of interference with host system environment.

*E. Service level agreements for cloud security*

Normally, Service Level Agreements (SLAs) intended to ensure 99.999

SLAs are normally published by cloud service providers. Clients can investigate and negotiate terms and conditions with service providers in order to ensure comprehensive security of system. In this way client is having much better overview of service providers security framework and able to choose PaaS services with more enhanced security infrastructure.

*F. Autonomic Security*

Autonomic security is a self-protection mechanism that auto-detects any harmful intrusion and take mitigation action according to the situation (Vieira et al., 2014). Autonomic security system offers self-configuration, self-awareness and self-healing capabilities to the cloud architecture. An autonomic system consists of best techniques of intrusion detection those offer excellent defense, possible in short time, against any outside attacks. This feature offers more time to system administrator to design new strategy and apply more security parameters to avoid future security vulnerabilities.

The basic objectives of implementing autonomic security are to keep the system-up and working in case of any security attack. Autonomic security ensures maximum availability and keeps the elements of cloud architecture according to design specification.

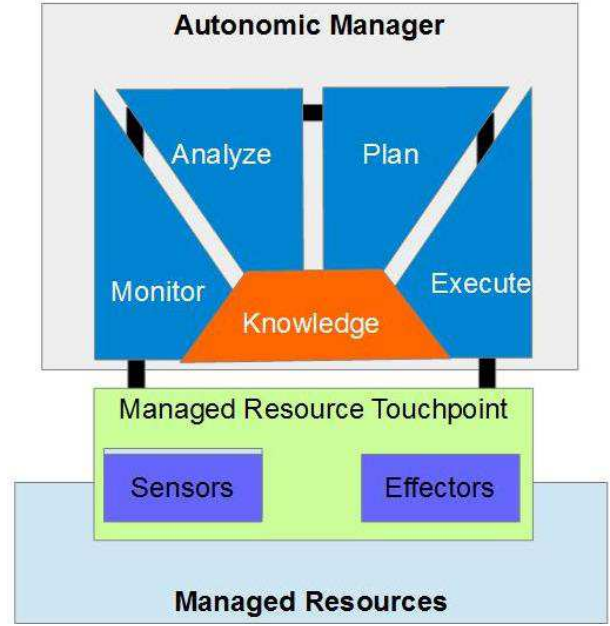

Figure 4. Autonomic Security module

Fig.4 depicts the structure of an autonomic system. It is based on the idea of an automated Intrusion Detection System (IDS) that have monitor, analysis, plan and execute modules (Wu et al., 2013). These all modules are managed by a meta-management element. This meta-management component is responsible for taking IDS based decisions through interpreting knowledge-base. In this module sensors are used to gather information from managed information resources.

## VII. RESULTS AND DISCUSSION

Implementation and analysis of a defined security model of MPSM are focused on functional specifications. We proposed an analysis of the model to ensure that every single security control and artifact is instantiated in a business security rule and is comprised in the conceptual model throughout the analysis phase. In order to assist this analysis we have used the reverse engineering tool Marble [1]. To identify and extract a number of security items in the proposed model, we have developed a series of templates from some of the different set of business rules that include the security goals, policies and requirements. After defining the actual security requirements for the cloud, we presented a number of cloud specific aspects like threats, security constraints and cloud service provider limitations. In this overall analysis, it is ensured that MPSM must comply all key security needs and ensure maximum security. A catalogue of security mechanisms has been developed based targeting key domain specific in the PaaS cloud security aspects. The key objective of this step is to ensure the security mechanism application against key threats. Secure Tropos [2] is a key tool that is employed to ensure the security requirements engineering methodology. Tropos stated guideline will be followed throughout the entire development process. Fig. 6 shows a graphical representation of the overall analysis activity of MPSM security application tasks together with the key input and output artifacts using SPEM 2.0 diagrams. The potential outcomes of the application of this model offer a great deal of security and efficiency as compared to traditional security practices. Currently, cloud computing adoption is growing; the demand for multitenant platforms is also increasing. Multitenant platforms are offering shared computing resources; therefore security management is a key challenge in such platforms. The present PaaS clouds design suffer from some significant security flaws. These flaws need to be addressed while using, adopting or designing PaaS systems. Figure.5 summarizes the key security issues and requirements for a PaaS platform. It also shows recommended mitigation model for aforementioned security issues. Fig. 7 has outlined the proposed security model that is offering a multi-perspective solution for mentioned security issues. This model is offering security management policies, tools and techniques for implementing a secure PaaS architecture. Besides implementing mentioned tools, security in PaaS clouds demands enforcement of security policies. Security policies are designed and issued by vendors while the client organizations implement those policies practically in order to ensure security from fundamental security risks. One of the key aspects of PaaS security policy is to restrict users not to develop apps that are themselves prone to security attack. Code monitoring is a strong security enforcement policy to avoid major security mishaps. Future PaaS development is moving toward the container based application development. Docker is one of the top vendors in this technology market. Container born apps are more fast and resource efficient than VM born apps. Though, container architecture is less secure than traditional VM based architecture. So, proposed future work is about the analysis of security vulnerabilities of container based PaaS systems and proposing security model to secure new generation of container based PaaS platforms.

---
[1] https://marketplace.eclipse.org/content/marble

| PaaS security Threats | PaaS Security Requirements | Mitigation Strategies |
|---|---|---|
| • Programming flaws<br>• Software modification<br>• Software interruption (deletion)<br>• Impersonation<br>• Session hijacking<br>• Traffic flow analysis<br>• Exposure in network<br>• Defacement<br>• Connection flooding<br>• DDOS<br>• Impersonation<br>• Disrupting communications | • Access control<br>• Application security<br>• Data security<br>• Cloud management control security<br>• Secure images<br>• Virtual cloud protection<br>• Communication security | *Trusted Cloud Computing*<br>• Trusted Hardware and software<br>• Encrypting devices<br>*Secure Execution and Communication Environment*<br>• Symmetric & Homomorphic encryption<br>• Secret socket layer<br>• Protection of data in transmission, storage and processing<br>*Identity and Access Management (IAM)*<br>• Authentication<br>• Authorization<br>• Traceability<br>*Virtualization Security Management*<br>• Harding the VM<br>• Harden the Hypervisor<br>• Firewall additional VM ports<br>• Root Securing Monitor<br>*Service level agreements for cloud security*<br>*Autonomic Security* |

Figure 5. Requirements, Threats and Mitigation

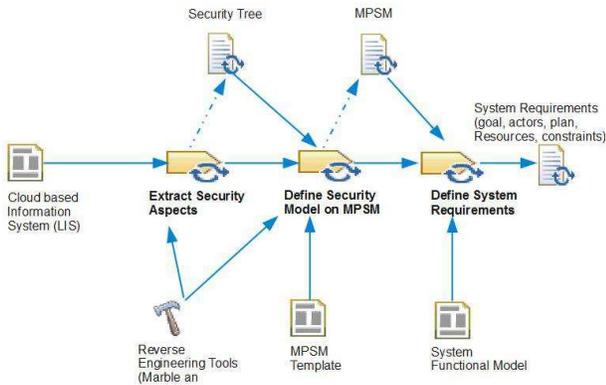

Figure 6. MPSM Implementation Flowchart

[2] http://www.troposproject.org/

## VIII. CONCLUSION

Secure PaaS platform is one of the key needs in current age. Currently, cost and time efficient application development is in high demand and PaaS is the ultimate technology to fulfill rising technology needs. This paper addressed security issues and vulnerabilities in PaaS cloud. A security model is proposed in this paper by considering the mentioned security vulnerabilities. This model offers a multi-perspective solution for implementing more secure PaaS infrastructure. Through implementing proposed model, PaaS platform would be resistant against the above mentioned security problems. Aforementioned mitigation strategies offer a deep and detailed insight for adopting a safe and secure PaaS platform. By knowing and adopting these solutions, organizations and customers both can be well-informed and precautionary while adopting and using PaaS platforms.

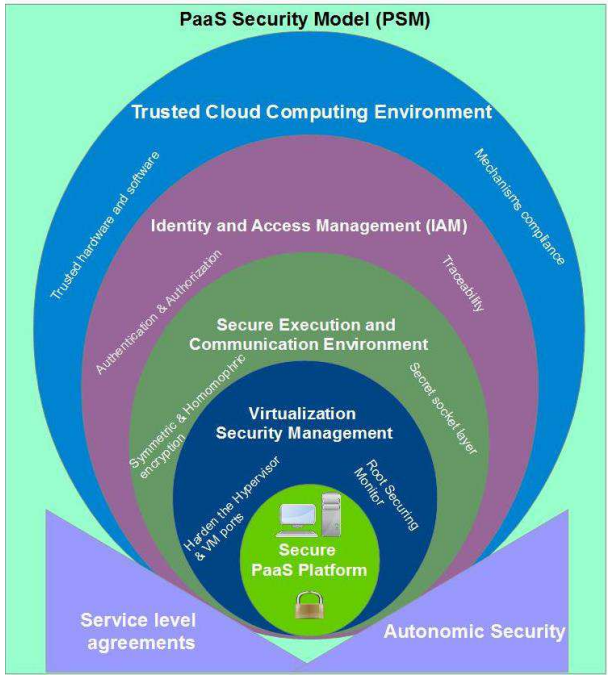

Figure 7. MPSM: A multi-perspective solution